\documentclass[cup6a]{cupbook}
\usepackage[dvips]{graphicx}

\begin{document}

\pagenumbering{roman}
\maketitle
\tableofcontents
\cleardoublepage
\pagenumbering{arabic}

\chapter{Supermassive Black Holes}

Supermassive black holes have generally been recognized as the most destructive 
force in nature. But in recent years, they have undergone a dramatic shift in 
paradigm. These objects may have been critical to the formation of structure in 
the early universe, spawning bursts of star formation and nucleating proto-galactic 
condensations. Possibly half of all the radiation produced after the Big Bang
may be attributed to them, whose number is now known to exceed 300 million. 
The most accessible among them is situated at the Center of Our Galaxy. In
the following pages, we will examine the evidence that has brought us to this 
point, and we will understand why many expect to actually image the event horizon
of the Galaxy's central black hole within this decade.

The supermassive black hole story begins in 1963, at the Mount Palomar observatory,
where Schmidt (1963) was pondering over the nature of a star-like object with truly 
anomalous characteristics.  Meanwhile, at the university of Texas, Kerr (1963) was 
making a breakthrough discovery of a solution to Einstein's field equations of
general relativity. Kerr's work would eventually produce a description of space 
and time surrounding a spinning black hole, now thought to power the compact 
condensations of matter responsible for producing the mystery on Schmidt's desk 
that year.

The development and use of radio telescopes in the 1940s had led to the gradual
realization that several regions of the sky are very bright emitters of 
centimeter-wavelength radiation. In the early 1960's, the British astronomer 
Cyril Hazard's idea of using lunar occultation to determine with which, if any, 
of the known visible astronomical objects the emitter of centimeter-wavelength 
radiation was associated, lead to the successful identification of 3C~273 as a 
star-like object in Virgo. Its redshifted lines, however, indicated that this was 
not a star at all, but rather an object lying at great cosmological distances.  
A recent image of this historic source was made with the {\sl Chandra} X-ray 
telescope, and is shown in figure~1.

\begin{figure}[h]
\centerline{\includegraphics[width=4in]{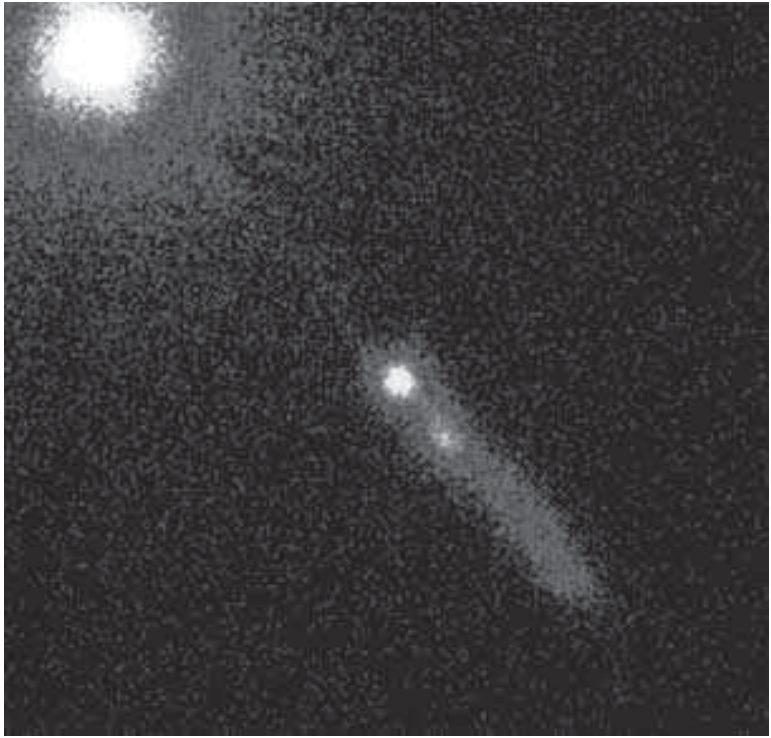}}
\vskip0.15in
\caption{The quasar 3C~273 (the bright object in the upper-left-hand corner) 
was one of the first objects to be recognized as a ``quasi-stellar-radio-source'' 
(quasar), due to its incredible optical and radio brightness. Insightful analysis
led to the realization that 3C~273 is actually an incredibly powerful, distant 
object.  This {\sl Chandra} image has a size $\approx 22\times 22$ square
arcseconds, which at the distance to 3C~273, corresponds to about $2,000\times 2,000$
square light-years.  (Photograph courtesy of H. L. Marshall et al., NASA, and MIT)}
\end{figure}

3C~273's total optical output varies significantly in only 10 months or so, implying 
that its size cannot exceed a few light-years---basically the distance between the 
Sun and its nearest stellar neighbor. So it was clear right from the beginning that 
the quasar phenomenon must be associated with highly compact objects. Even more
impressively, their X-ray output has now been seen to vary in a matter of only hours,
corresponding to a source size smaller than Neptune's orbit.  Each quasar typically 
releases far more energy than an entire galaxy, yet the central engine that drives 
this powerful activity occupies a region smaller than our solar system.

The idea that such small volumes could be producing the power of a hundred billion
Suns led to their early identification as radiative manifestations of supermassive
black holes (see, e.g., Salpeter 1964, Zel'dovich and Novikov 1967, and
Lynden-Bell 1969).  But are they ``naked''---deep, dark pits of matter floating 
aimlessly across the primeval cosmic soup---or are they attached to more 
recognizable structures in the early universe?

In recent years, the task of source identification has been made easier using the Hubble
Space Telescope (see figure~2).  The most widely accepted view now is that quasars 
are found in galaxies with active, supermassive black holes at their centers. 
Because of their enormous distance from Earth, the ``host"  galaxies appear very 
small and faint, and are very hard to see against the much brighter quasar light 
at their center.  

\vskip 0.1in
\begin{figure}[h]
\centerline{\includegraphics[width=4in]{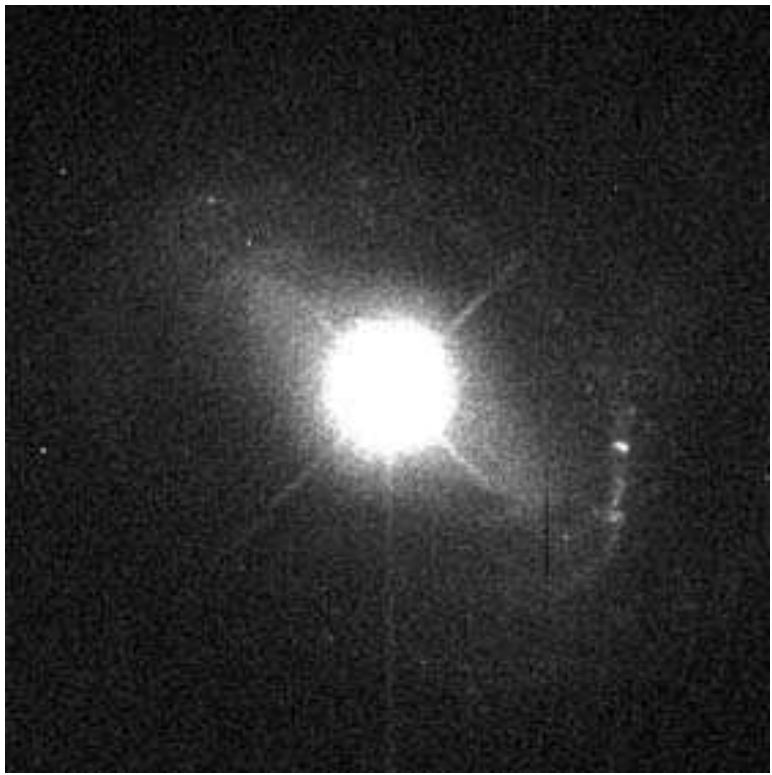}}
\vskip -0.0in
\caption{This Hubble Space Telescope (HST) image reveals the faint host galaxy within
which dwells the bright quasar known as QSO 1229+204. The quasar is seen to lie in the 
core of an ordinary-looking galaxy with two spiral arms of stars connected by a bar-like
feature.  (Photograph courtesy of John Hutchings, Dominion Astrophysical Observatory, 
and NASA)}
\end{figure}

Quasars actually reside in the nuclei of many different types of galaxy, from 
the normal to those highly disturbed by collisions or mergers.  A supermassive 
black hole at the nucleus of one of these distant galaxies ``turns on" when it 
begins to accrete stars and gas from its nearby environment; the rate at which 
matter is converted into energy can be as high as 10 solar masses per year.  So 
the character and power of a quasar depend in part on how much matter is available
for consumption.  Disturbances induced by gravitational interactions with
neighboring galaxies can trigger the infall of material toward the center of the
quasar host galaxy (see figure~3).  However, many quasars reside in apparently 
undisturbed galaxies, and this may be an indication that mechanisms other than 
a disruptive collision may also be able to effectively fuel the supermassive 
black hole residing at the core.

\begin{figure}[h]
\centerline{\includegraphics[width=4in]{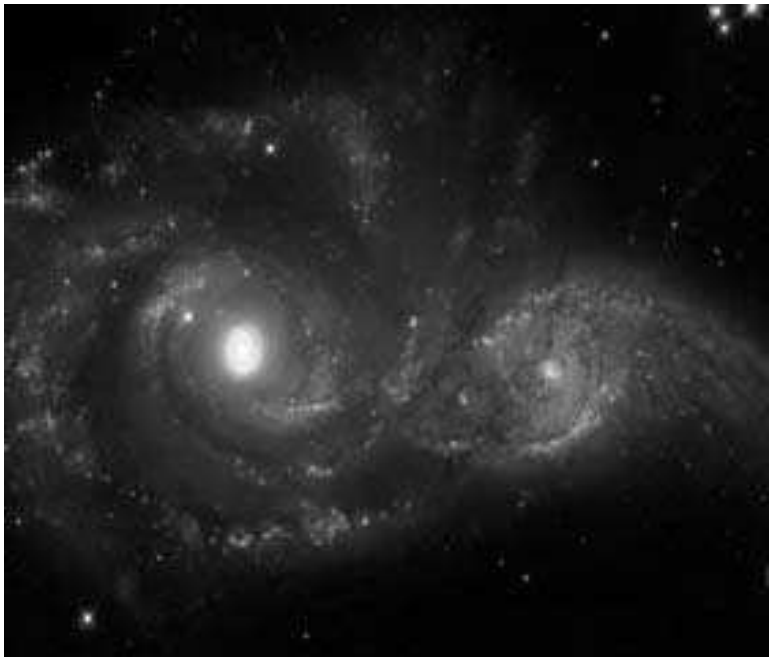}}
\vskip 0.15in
\caption{The collision between two galaxies begins with the unraveling of
the spiral disks.  This HST image shows the interacting pair of galaxies NGC~2207 (the
larger, more massive object on the left) and IC~2163 (the smaller one on the right),
located some 114 million light-years from Earth.  By this time, IC~2163's stars have
begun to surf outward to the right on a tidal tail created by NGC~2207's strong
gravity. (Image courtesy of Debra M. Elmegreen at Vassar College, Bruce G. Elmegreen
at the IBM Research Division, NASA, M. Kaufman at Ohio State U., E. Brinks at U. de
Guanajuato, C. Struck at Iowa State U., M.  Klaric at Bell South, M. Thomasson at
Onsala Space Observatory, and the Hubble Heritage Team based at the Space Telescope
Space Institute and AURA)}
\end{figure}

Some supermassive black holes may not be visible as quasars at all, but rather just
sputter enough to become the fainter galactic nuclei in our galactic neighborhood.
Our Milky Way galaxy and our neighbor, the Andromeda galaxy, harbor supermassive 
black holes with very little nearby plasma to absorb.  The question concerning how 
the undisturbed galaxies spawn a quasar is still not fully answered.  Perhaps the 
Next Generation Space Telescope, now under development and expected to fly soon
after 2010, will be able to probe even deeper than the Hubble Space Telescope has 
done, and expose the additional clues we need to resolve this puzzle.

\section{The Quasar/Supermassive Black Hole Census}
By now, some 15,000 distant quasars have been found, though the actual number of 
supermassive black holes discovered thus far is much greater.  Because of their 
intrinsic brightness, the most distant quasars are seen at a time when the universe 
was a mere fraction of its present age, roughly one billion years after the Big Bang.  
The current distance record is held by an object found with the Sloan Digital Sky 
Survey (SDSS), with a redshift of $\sim 6.3$, corresponding to a time roughly 700 
million years after the Big Bang.

The SDSS has shown that the number of quasars rose dramatically from a billion years 
after the Big Bang to a peak around 2.5 billion years later, falling off sharply at 
later times toward the present.  Quasars turn on when fresh matter is brought into 
their vicinity, and then fade into a barely perceptible glimmer not long thereafter. 

However, not all the supermassive black holes in our midst have necessarily
grown through the quasar phase. Quasars typically have masses $\sim 10^9\;
M_\odot$. Yet the black hole at the center of our galaxy is barely $3.4\times
10^6\;M_\odot$. In other words, not all the supermassive black holes in our 
vicinity are dormant quasars. 

A recent discovery suggests how some of these ``smaller'' black holes may
have gotten their start. {\sl Chandra} has identified what appears to be a 
mid-sized black hole 600 light-years from the center of M82 (Matsumoto et al. 
2001).  With a mass of $\sim 500\;M_\odot$, this object could conceivably 
sink to the center of M82, and then grow to become a supermassive black hole 
in its own right, without having passed through the rapid accretion phase of
a quasar.

So the class of known quasars may be a good tracer of supermassive black holes,
but it clearly does not encompass all of them. Taking advantage of two patches
of sky relatively devoid of nearby objects, {\sl Chandra} produced two of the 
deepest images ever made of the distant cosmos at X-ray energies, one in the 
southern hemisphere and the other in the north---the latter, called the 
{\sl Chandra} Deep Field North, is shown in figure~4. Based on the number of
suspected supermassive black holes in these images, one infers an overall
population of $\sim 300$ million throughout the cosmos.

\begin{figure}[h]
\centerline{\includegraphics[width=4in]{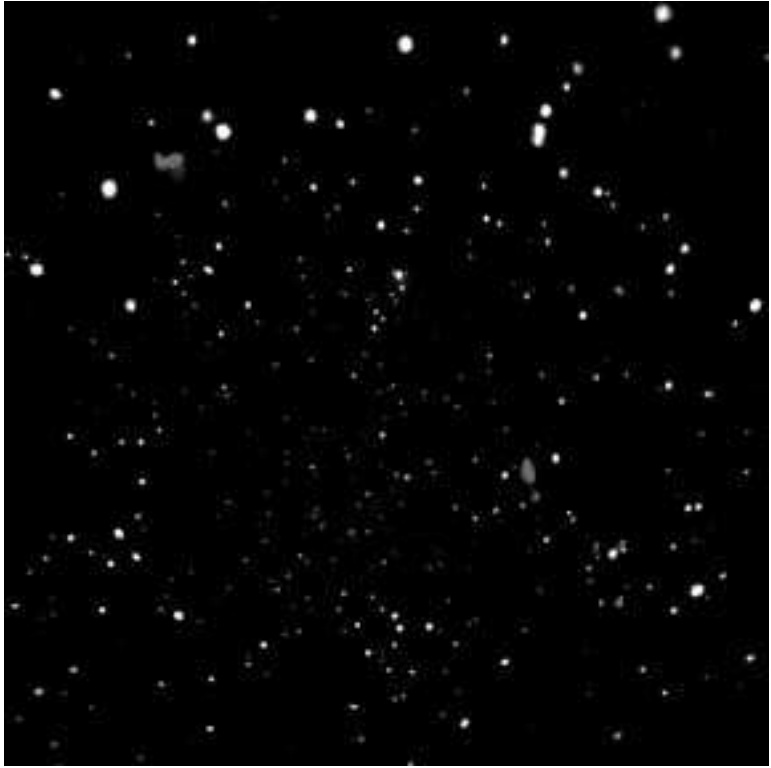}}
\vskip -0.0in
\caption{The {\sl Chandra} Deep Field North X-ray image. The vast majority of 
the 500 or so sources in this view (spanning a region approximately 28 
arcminutes wide) are supermassive black holes.  Extrapolating this number to 
the whole sky, one infers $\sim 300$ million such objects spread across the 
universe. (Image courtesy of D. M. Alexander, F. E. Bauer, W. N. Brandt, 
et al., NASA and PSU)}
\vskip 0.0in
\end{figure}

And yet, these X-ray detections speak only of those particular supermassive 
black holes whose orientation facilitates the transmission of their high-energy 
radiation toward Earth. Their actual number must be higher than this;  indeed, 
there is now growing evidence that many---perhaps the majority---of the 
supermassive black holes in the universe are obscured from view.  The faint 
X-ray background pervading the intergalactic medium has been a puzzle for many years.
Unlike the cosmic microwave background radiation left over from the Big Bang, the 
photons in the X-ray haze are too energetic to have been produced at early times.  
Instead, this radiation field suggests a more recent provenance associated with 
a population of sources whose overall radiative output may actually dominate over 
everything else in the cosmos.  Stars and ordinary galaxies simply do not radiate 
profusely at such high energy, and therefore cannot fit the suggested profile.

A simple census shows that in order to produce such an X-ray glow with quasars alone, 
for every known source there ought to be ten more obscured ones.  This would also mean 
that the growth of most supermassive black holes by accretion is hidden from the view 
of optical, UV, and near infrared telescopes.  Fabian et al. (2000) have reported the 
discovery of an object they call a Type-2 quasar.  Invisible to optical light telescopes, 
the nucleus of this otherwise normal looking galaxy betrayed its supermassive guest with 
a glimmer of X-rays.  The implication is that many more quasars, and their supermassive 
black-hole power sources, may be hidden in otherwise innocuous-looking galaxies.

And so, the all-pervasive X-ray haze, in combination with the discovery of gas-obscured 
quasars, now point to supermassive black holes as the agents behind perhaps {\it half} 
of all the universe's radiation produced after the Big Bang. Ordinary stars no longer 
monopolize the power as they had for decades prior to the advent of space-based astronomy.

\section{Black Holes in the Nuclei of Normal Galaxies}
Much closer to Earth, within hundreds of thousands of light-years as opposed to the 
11 billion-light-year distance to the farthest quasars, supermassive black holes 
accrete at a lower rate than their quasar brethren and are therefore much fainter. 
An archetype of this group, Centaurus A, graces the southern constellation of Centaurus
at a distance of 11 million light-years (see figure~5). At the center of the dark bands of dust,
HST recently uncovered a disk of glowing, high-speed gas, swirling about a concentration of matter
with the mass of $\sim 2\times 10^6\;M_\odot$.  This enormous mass within the central cavity 
cannot be due to normal stars, since these objects would shine brightly, producing an intense 
optical spike toward the middle, unlike the rather tempered look of the infrared image shown here.

\begin{figure}[h]
\centerline{\includegraphics[width=4in]{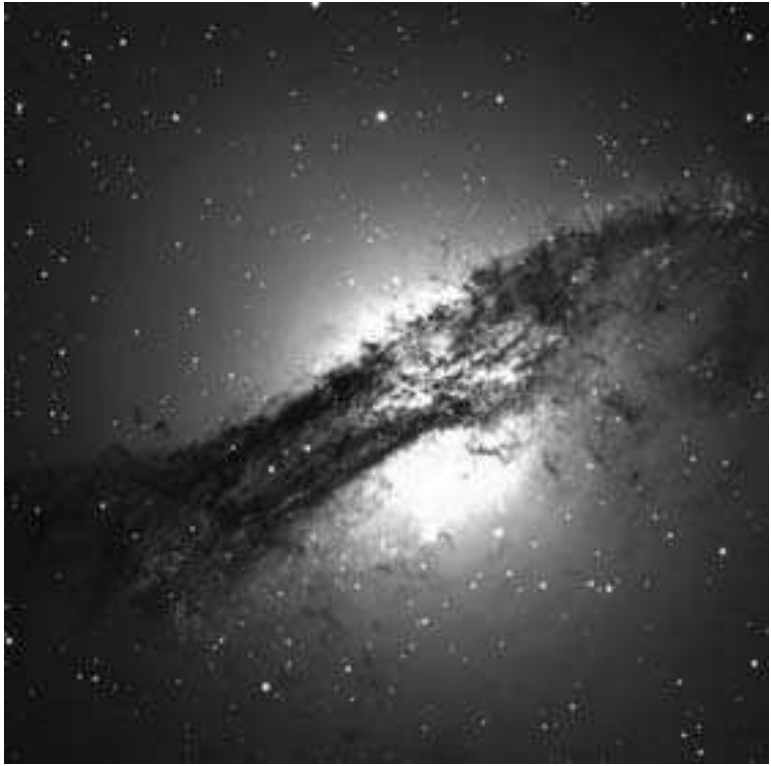}}
\vskip -0.0in
\caption{This image of Centaurus A is a composite of three photographs taken by the 
European Southern Observatory. The dramatic dark band is thought to be the remnant 
of a smaller spiral galaxy that collided, and ultimately merged, with a large 
elliptical galaxy.  Imaging this galaxy at radio wavelengths, we would see two 
jets of plasma spewing forth from the central region in a direction perpendicular 
to the dark dust lanes (see figure~6 for the corresponding configuration in
Cygnus A).  These relativistic expulsions of plasma share much in common with the
X-ray glowing stream shown in figure~1. (Photograph courtesy of Richard M. West and the
European Southern Observatory)}
\end{figure}

Centaurus A is apparently funneling highly energetic particles into beams perpendicular
to the dark strands of dust. It may therefore have much in common with the X-ray
jet-producing black hole in 3C 273 (see figure~1), and another well-known active
galactic nucleus, Gygnus A, shown in figure~6.  The luminous extensions in this figure
project out from the nucleus of Cygnus A, an incredible distance three times the size of
the Milky Way.  Yet located 600 million light-years from Earth, they cast an aspect
only one-tenth the diameter of the full moon. Radio and X-ray observations show that 
objects such as this accelerate plasma to relativistic speeds on scales of 10 to 100 
Schwarzschild radii (see figure~7), and that these jets are not rare.

\begin{figure}[h]
\centerline{\includegraphics[width=\textwidth]{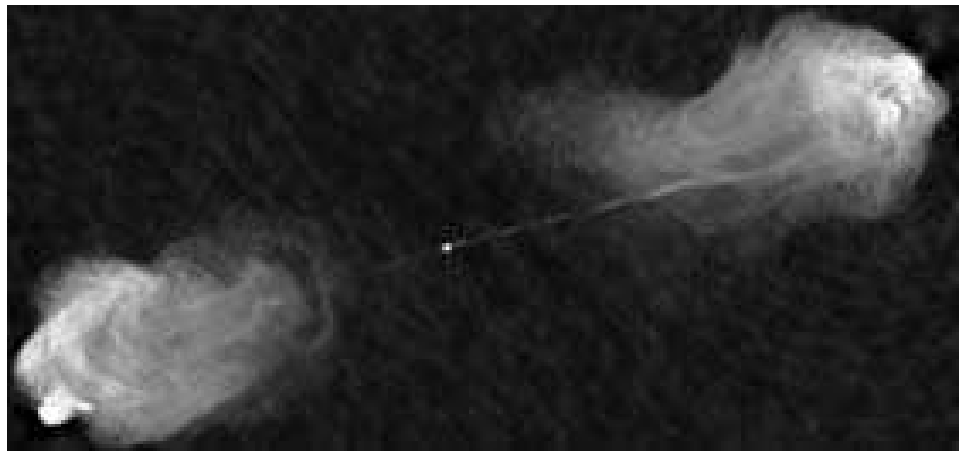}}
\vskip -0.0in
\caption{A VLA image of the powerful central engine and its relativistic ejection of plasma 
in the nucleus of Cygnus A.  Taken at $6$ cm, this view reveals the highly ordered structure 
spanning over 500,000 light-years, fed by ultra-thin jets of energetic particles beamed from 
the compact radio core between them. The giant lobes are themselves formed when these jets 
plow into the tenuous gas that exists between galaxies.  Despite its great distance from us 
(over 600 million light-years), it is still by far the closest powerful radio galaxy and one 
of the brightest radio sources in the sky. The fact that the jets must have been sustained 
in their tight configuration for over half a million (possibly as long as ten million) 
years means that a highly stable central object---probably a rapidly spinning supermassive 
black hole acting like an immovable gyroscope---must be the cause of all this activity.  
(Photograph courtesy of Chris Carilli and Rick Perley,
NRAO, and AUI)}
\end{figure}

An important conclusion to draw from the morphology of jets like those in Cygnus A is that
the process responsible for their formation must be stable for at least as long as it takes 
the streaming particles to journey from the center of the galaxy to the extremities of the 
giant radio lobes. Evidently, these pencil-thin jets of relativistic plasma have retained 
their current configuration for over one million years.  The most conservative view regarding
the nature of these objects is that a spinning black hole is ultimately responsible for this
activity. The axis of its spin functions as a gyroscope, whose direction determines the 
orientation of the jets.  Although the definitive mechanism for how the ejection takes place 
is yet to be determined, almost certainly the twisting motion of magnetized plasma near the 
black hole's event horizon is causing the expulsion.  The Kerr spacetime, which describes the 
dragging of inertial frames about the black hole's spin axis, provides a natural setting for 
establishing the preferred direction for this ejective process.

\begin{figure}[h]
\centerline{\includegraphics[width=\textwidth]{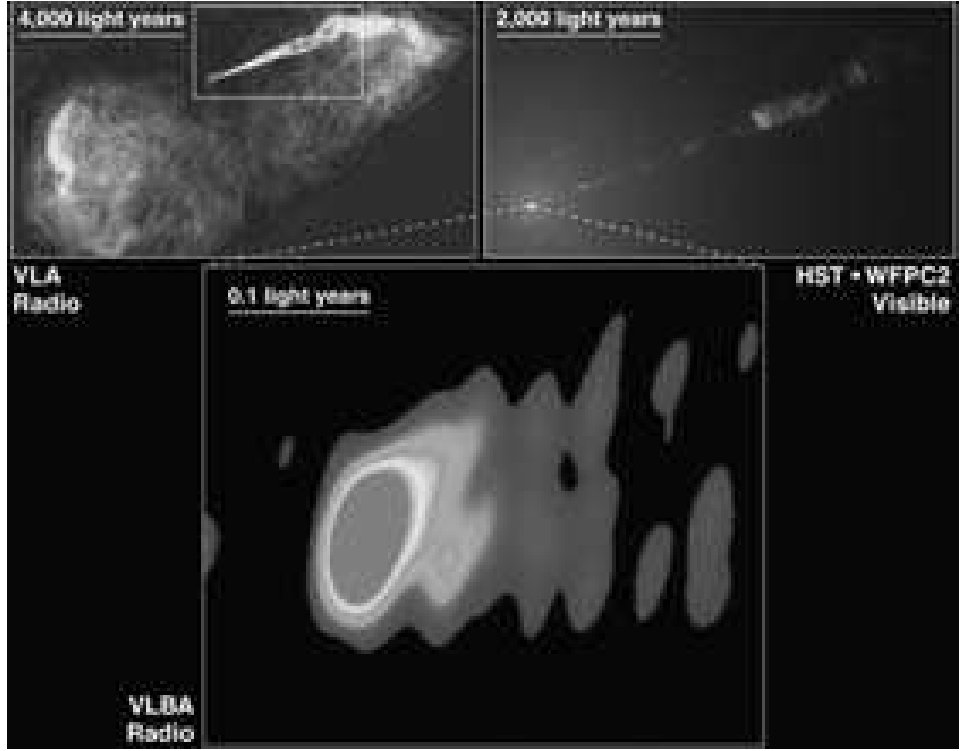}}
\vskip -0.05in
\caption{Streaming out from the center of M87 is a black-hole-powered jet of 
plasma traveling at near lightspeed.  Its source is a powerful central object 
with a mass of $\sim 3\times 10^9\;M_\odot$.  This sequence of photographs shows 
progressively magnified views: {\sl top left:} a VLA image showing the full extent 
of the jets and the blobs at the termination points; {\sl top right:} a visible 
light image of the giant elliptical galaxy M87, taken with NASA's Hubble Space 
Telescope;  {\sl bottom:} a Very Long Baseline Array (VLBA) radio image of the 
region surrounding the black hole.  This view shows how the extragalactic jet 
is formed into a narrow beam within a few tenths of a light-year of the nucleus 
(the red region is only a tenth of a light-year across), corresponding to only
100 Schwarzschild radii for a black hole of this mass. (Photographs courtesy of
the National Radio Astronomy Observatory and the National Science Foundation
[top left and bottom], and John Biretta at the Space Telescope Science Institute,
and NASA [top right])}
\end{figure}

\section{Weighing Supermassive Black Holes}
Black hole masses are measured with a variety of techniques, though all have to
do with the dynamics of matter within their gravitational influence. Knowledge
of the radiating plasma's distance from the central object, and the force required 
to sustain its motion at that distance, is sufficient for us to extract the 
central mass.

One of the more compelling applications of this technique has been made to the 
spiral galaxy NGC~4258. Using global radio interferometry, Miyoshi et al. (1995) 
observed a disk of dense molecular material orbiting within the galaxy's nucleus 
at speeds of up to 650 miles per second. This disk produces sufficient radiation 
to excite condensations of water molecules, leading to strong maser emission at 
radio wavelengths. The disk within which these water molecules are trapped is
small compared to the galaxy itself, but it happens to be oriented fortuitously 
so that beams of microwaves are directed along our line-of-sight.

The maser clouds appear to trace a very thin disk, with a motion that follows 
Kepler's laws to within one part in 100, reaching a velocity (inferred from
the Doppler shift of the lines) of about 650 miles per second at a distance of 0.5 
light-years from the center.  The implied central mass is $\sim (35-40)\times
10^6\;M_\odot$, concentrated within 0.5 light years of the center in NGC~4258.
This points to a matter density of at least $10^8\;M_\odot$ per cubic light-year.  
If this mass were simply a highly concentrated star cluster, the stars would be 
separated by an average distance only somewhat greater than the diameter of the 
solar system, and with such proximity, they would not be able to survive the 
inevitable catastrophic collisions and mergers with each other. Because of the 
precision with which we can measure this concentration of dark mass, we regard 
the object in the nucleus of NGC~4258 as one of the two most compelling 
supermassive black holes now known, the other being the object sitting at the 
center of our own galaxy, about which we will have more to say shortly.

The fortuitous arrangement of factors that permits the use of this particular 
technique does not occur often, however, so other methods must be used. 
Often, clouds of gas orbiting the nucleus are irradiated by the central engine, 
and they in turn produce a spectrum with emission lines indicative of the plasma's 
ionized state.  The method used to determine the distance of these ionized clouds 
from the black hole is known as reverberation.  By monitoring the light emitted by 
the supermassive black hole and, independently, the radiation from its halo of 
irradiated clouds of gas, one can determine when a variation in the radiative 
output has occurred.  When the quasar varies its brightness, so does the 
surrounding matter---but only after a certain time delay.  The lag is clearly 
due to the time it took the irradiating light from the center to reach the clouds, 
and using the speed of light, this delay provides a measure of the distance between 
the nucleus and the orbiting plasma.  Again, this procedure provides the speed of
matter and its distance from the center, which together yield a determination of
the gravitating mass.

Having said this, the best mass determination one can make is still based on kinematic
studies of stars orbiting the central object. The center of our Galaxy is close enough
for this method to work spectacularly.  Known as Sagittarius A*, the black hole
at the center of the Milky Way may not be the most massive, nor the most energetic, 
but it is by far the closest, only 28,000 light-years away. Figure~10 shows an infrared 
image of Galactic center produced recently with the 8.2-meter VLT YEPUN telescope at the
European Southern Observatory in Paranal, Chile.  The image we see here is sharp because 
of the use of adaptive optics, in which a mirror in the telescope moves constantly to 
correct for the effects of turbulence in the Earth's atmosphere.  Sagittarius A* is so 
close to us compared to other supermassive black holes, that on an image such as this,
we can identify individual stars orbiting a mere seven to 10 light-days from the source 
of gravity. In the nucleus of Andromeda, the nearest major galaxy to the Milky Way, the 
best we could do right now is about two light-years.

\begin{figure}[h]
\centerline{\includegraphics[width=4in]{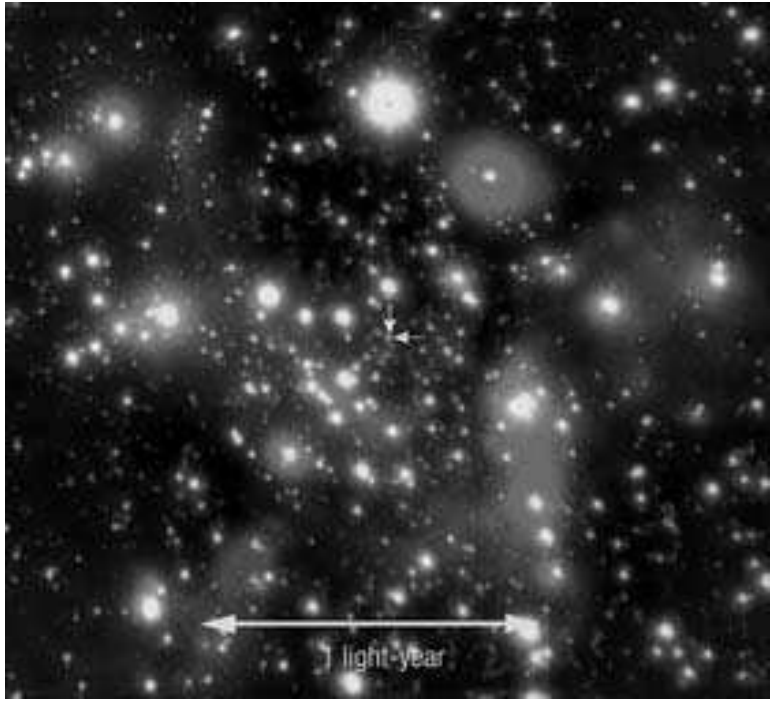}}
\vskip 0.2in
\caption{This ($1.6-3.5$ micron) image, taken by the European Southern Observatory's 
8.2-meter telescope atop Paranal, Chile, provides one of the sharpest views of the 
stars surrounding the supermassive black hole at the Galactic center.  The location 
of the black hole itself is indicated by the two central arrows.  This view represents 
a scale of approximately $2\times 2$ square light-years.  (Photograph courtesy of R. 
Genzel et al. at the Max-Planck-Institut f\"ur Extraterrestrische Physik, and the 
European Southern Observatory)}
\end{figure}

In the Galactic center, stars orbit Sagittarius A* at speeds of up to five million 
kilometers per hour. This motion is rapid, in fact, that we can easily detect their
proper motion on photographic plates taken only several years apart. Some of them
complete an orbit about the center in only 15 years (Sch\"odel, Ott, Genzel, et al. 
2002).  In the middle of the photograph in figure~8, it appears that one of the 
fainter stars---designated as S2---lies right on top of the position where the 
black hole is inferred to be.  S2 is an otherwise ``normal" star, though some 
15 times more massive and seven times larger than the Sun. This star, S2, has
now been tracked for over ten years and the loci defining its path trace a perfect 
ellipse with one focus at the position of the supermassive black hole.  This 
photograph, taken near the middle of 2002, just happens to have caught S2 at the 
point of closest approach (known as the perenigricon), making it look like it was
sitting right on top of the nucleus.

At perenigricon, S2 was a mere 17 light-hours away from the black hole---roughly 
three times the distance between the Sun and Pluto, while traveling with a speed 
in excess of 5,000 kilometers per second, by far the most extreme measurements
ever made for such an orbit and velocity. We infer from these data that the mass
of Sagittarius A* is $\sim 3.4\times 10^6\;M_\odot$, compressed into a region no 
bigger than $\sim 17$ light-hours.  For this reason, Sagittarius A*, and the
central object in NGC~4258, are considered to be the most precisely ``weighed'' 
supermassive black holes yet discovered.

\section{The Formation of Supermassive Black Holes}
An increasingly important question being asked in the context of supermassive black holes
is which came first, the central black hole, or the surrounding galaxy?  Quasars seem
to have peaked 10 billion years ago, early in the universe's existence.  The light from 
galaxies, on the other hand, originated much later---after the cosmos had aged another 
2 to 4 billion years. Unfortunately, both measurements are subject to uncertainty, and 
no one can be sure we are measuring {\it all} of the light from quasars and galaxies, so 
this argument is not quite compelling.  But we do see quasars as far out as we can look,
and the most distant among them tend to be the most energetic objects in the
universe, so at least {\it some} supermassive black holes must have existed near
the very beginning.  At the same time, images such as figure~3 provide evidence of mergers 
of smaller structures into bigger aggregates, but without a quasar.  Perhaps not every 
collision feeds a black hole or, what is more likely, at least some galaxies must have 
formed first. Several scenarios for the formation of supermassive black holes are currently 
being examined.  In every case, growth occurs when matter condenses following either the 
collapse of massive gas clouds, or the catabolism of smaller black holes in collisions and mergers.

All of the structure in the universe traces its beginnings to a brief era
shortly after the Big Bang. Very few ``fossils'' remain from this period; one
of the most important is the cosmic microwave background radiation.
The rapid expansion that ensued lowered the matter density and temperature,
and about one month after the Big Bang, the rate at which photons were created
and annihilated could no longer keep up with the thinning plasma. The radiation
and matter began to fall out of equilibrium with each other, forever imprinting
the conditions of that era onto the radiation that reaches us to this day from all
directions in space.

We now know that the temperature anisotropies are smaller than one part in
a thousand, a limit below which density perturbations associated with
ordinary matter would not have had sufficient time to evolve freely
into the nonlinear structures we see today. Only a gravitationally
dominant dark-matter component could then account for the strong
condensation of mass into galaxies and supermassive black holes.
The thinking behind this is that whereas the cosmic microwave
background radiation interacted with ordinary matter, it would
retain no imprint at all of the dark matter constituents in the
universe.  The nonluminous material could therefore be condensed
unevenly (sometimes said to be `clumped') all the way back to the
Big Bang and we simply wouldn't know it.

The first billion years of evolution following the Big Bang must have been
quite dramatic in terms of which constituents in the universe would eventually
gain primacy and lasting influence on the structure we see today.  The issue of how
the fluctuations in density, mirrored by the uneven cosmic microwave background radiation,
eventually condensed into supermassive black holes and galaxies is currently
a topic of ongoing work.  This question deals with the fundamental contents of 
the universe, and possibly what produced the Big Bang and what came before it. 
The evidence now seems to be pointing to a coeval history for these two dominant 
classes of objects---supermassive black holes and galaxies---though as we have 
already noted, at least some of the former must have existed quite early.

One possibility proposed by Balberg and Shapiro (2002) is that the first 
supermassive objects formed from the condensation of dark matter alone; only
later would these seed black holes have imposed their influence on the latter.
But this dark matter has to be somewhat peculiar, in the sense that its
constituents must be able to exchange heat with each other.  As long as this
happens, a fraction of its elements evaporate away from the condensation,
carrying with them the bulk of the energy, and the rest collapse and create
an event horizon.  The net result is that the inner core of such a clump forms
a black hole, leaving the outer region and the extended halo in equilibrium
about the central object.  Over time, ordinary matter gathers around it,
eventually forming stars, and planets.

Ordinary matter could not have achieved this early condensation because
it simply wasn't sufficiently clumped initially.  Perhaps this material 
formed the first stars, followed by more stars, eventually assembling a 
cluster of colliding objects.  Over time, the inner core of such an assembly 
would have collapsed due to the evaporation of some of its members and the 
ensuing loss of energy into the extended halo, just as the dark matter did 
(see, e.g., Haehnelt and Rees 1993).  A seed black hole might have formed
in the cluster's core. Estimates show that once formed, such an object 
could have doubled its mass every 40 million years, so over the age of the 
universe, even a modestly appointed black hole could have grown into a 
billion-solar-mass object. The problem is that this could not have happened
in only 700 million years, when the first supermassive black holes appeared.

Yet another method leading to black hole growth results from ongoing
collisions between galaxies, which eventually lead to the merger of the
black holes themselves. An example of such a process occurring right now 
is shown in figure~9. 

\begin{figure}[h]
\centerline{\includegraphics[width=4in]{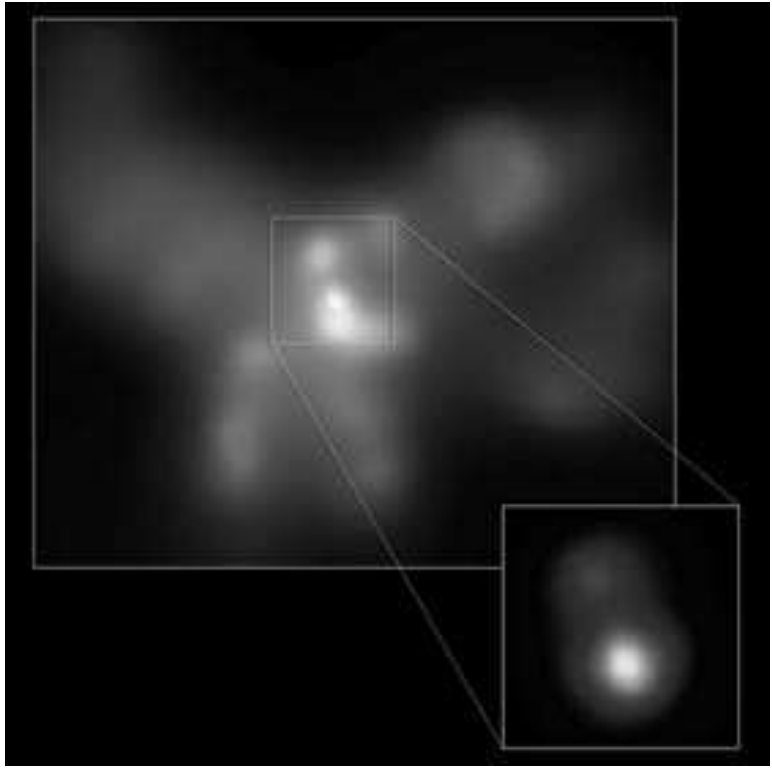}}
\vskip -0.0in
\caption{NGC~6240 is a butterfly-shaped galaxy, believed to be the
product of a collision between two smaller galaxies some 30 million
years ago.  This {\sl Chandra} X-ray image (34,000 light-years across)
shows the heat generated by the merger activity, which created the extensive
multimillion degree Celsius gas. We see here for the first time direct
evidence that the nucleus of such a structure contains not one, but
two active supermassive black holes, drifting toward each other across 
their 3,000-light-year separation; they're expected to merge into a bigger
object several hundred million years hence. (Image courtesy of Susan Komossa, 
Gunter Hasinger, and Joan Centrella, and the Max-Planck-Institut f\"ur 
Extraterrestrische Physik and NASA)}
\end{figure}

Almost every large, normal galaxy harbors a supermassive black hole at its 
center. Some evidence for this has been provided by a recently completed survey
of 100 nearby galaxies using the VLA, followed by closer scrutiny
with the Very Long Baseline Interferometry array. At least 30 percent of
this sample showed tiny, compact central radio sources bearing the
unique signature of the quasar phenomenon. Of the hundreds of millions
of supermassive black holes seen to pervade the cosmos, none of them
appear to be isolated. And even more compelling is the work of Kormendy 
and Richstone (1995), who set about the task of systematically measuring as 
many black hole masses as is currently feasible.  Direct measurements of 
supermassive black holes have been made in over 38 galaxies, based on the 
large rotation and random velocities of stars and gas near their centers.  
These objects are all relatively nearby because these direct methods don't 
work unless we can see the individual stars in motion about the central 
source of gravity. Curiously, none of the supermassive black holes have 
been found in galaxies that lack a central bulge (figure~10).  Galaxies 
with a central bulge may have undergone one or more mergers in their past. 
Thus, a collision like that seen in figures 3 and 11 may have been
required to create a central supermassive object. 

\begin{figure}[h]
\centerline{\includegraphics[width=4in]{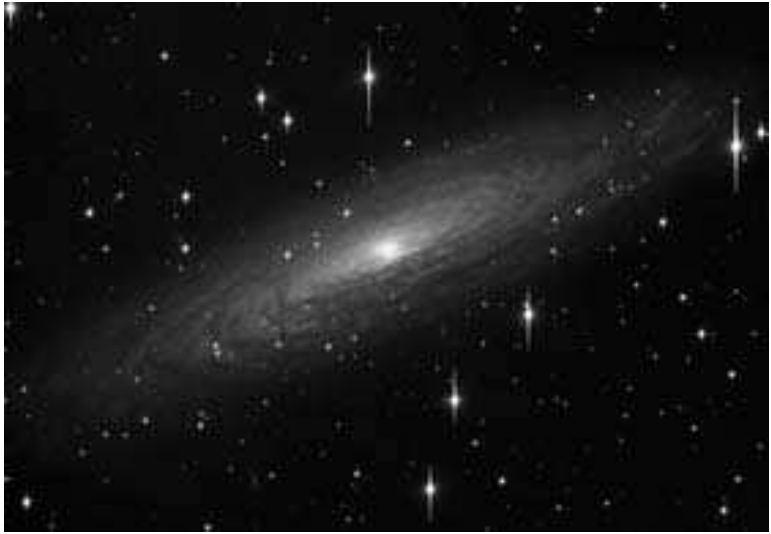}}
\vskip -0.0in
\caption{This image of the spiral galaxy NGC~2613 was captured on February 26, 2002, by the Very
Large Telescope in Paranal, Chile.  A search for a supermassive black hole in its core produced
a null result (Bower et al. 1993).  By now the preponderance of evidence suggests that highly
flattened disk galaxies lacking a significant central hub, or spheroidal component, also lack
a supermassive object in the nucleus.  On the other hand, every galaxy that does possess
a central bulge also harbors a supermassive black hole. (Image courtesy of S. D'Odorico et al.,
who obtained it during the test phase of the VIMOS instrument at the European Southern Observatory's
8-meter MELIPAL telescope, and courtesy also of the European Southern
Observatory)}
\end{figure}

Another recently inferred correspondence between supermassive black holes and their 
host galaxies seems to have clinched the case for a coeval growth (Ferrarese and Merritt 
2000; Gebhardt et al. 2000).  The data displayed in figure~11 show the relationship 
between the mass of the black hole and the velocity dispersion of stars within the 
spheroidal (or bulge) component of stars in the host galaxy. Based on the very tight 
correlation evident in this graph, it appears that the mass of the central black hole
can be predicted with remarkable accuracy simply by knowing the velocity of stars 
orbiting so far from it that its gravitational influence could not possibly be affecting 
their motion at the present time. Pinpointing the trajectory of a single object in the 
nuclei of the host galaxies is difficult and impracticable.  Instead, these measurements 
are based on the accumulated light from a limited region of the galaxy's central hub,
from which one may extract the collective Doppler shift, and thereby an average speed 
of the stars as a group. What these studies reveal is that the ratio of the black hole's 
mass to this average speed is constant across the whole sample of galaxies surveyed.

\begin{figure}[h]
\centerline{\includegraphics[width=\textwidth]{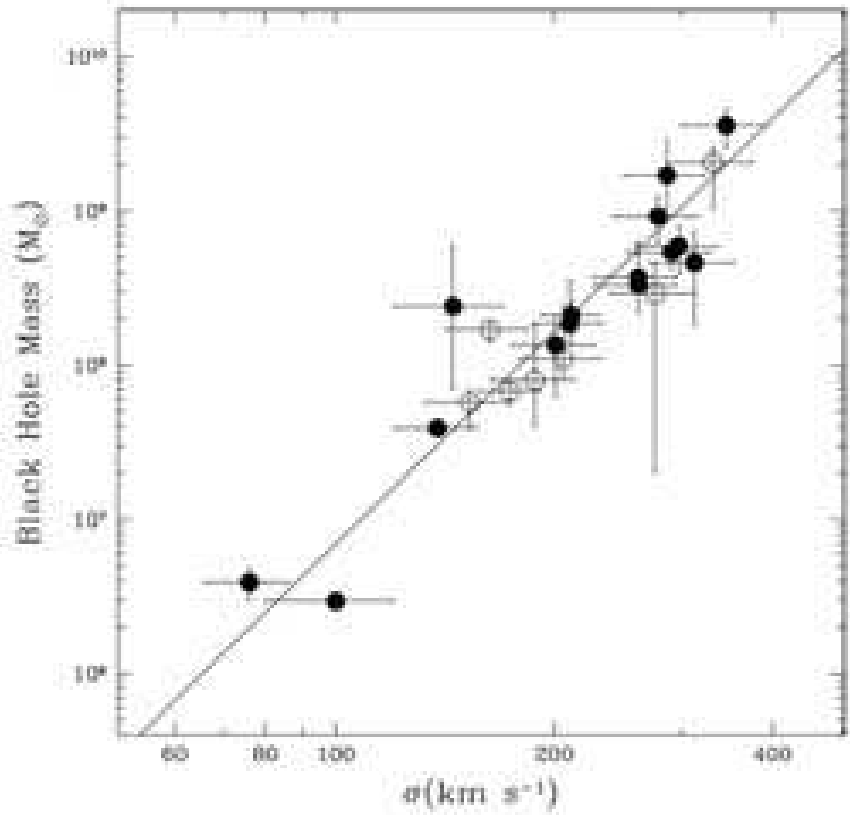}}
\vskip -0.1in
\caption{Plotted against the velocity dispersion of stars orbiting within the
spheroidal component of the host galaxy, the mass of a supermassive black hole
is tightly correlated with the former, hinting at a coeval history.
The stars are simply too far away from the nucleus to be directly affected by
the black hole's gravity at the present time (Ferrarese and Merritt 2000).}
\vskip 0.0in
\end{figure}

This result is one of the most surprising correlations yet discovered in the 
study of how the universe acquired its structure.  Supermassive black holes, 
it seems, ``know" about the motion of stars that are too distant to directly feel
their gravity.  This tight connection suggests an entangled history between a 
central black hole and the stellar activity in its halo. Although they may not 
be causally bonded today, they must have had an overlapping genesis in the past.

Additional information on the possible coeval history of supermassive black holes 
and their host galaxies is provided by images such as figure~4.  If we adopt a 
canonical efficiency of $10\;\%$ for the conversion of rest mass energy into 
radiation, we obtain the mass inflow rate versus redshift relation shown in 
figure~12.  This rate was as high as $10$ solar masses per year at a redshift 
of $3$ or so, and has declined monotonically to its much lower value of around 
$0.01$ solar masses per year in the current epoch.  An even more interesting 
evolutionary trait is that obtained by integrating the data in figure~12 over 
the black hole spatial distribution, which produces the redshift dependence 
of the so-called accretion rate density (figure~13).  This quantity effectively 
gives us the rate at which black hole mass is increasing with time.  The apparent 
$(1+z)^3$ dependence of this quantity (exhibited in figure~13) is identical to the
universe's star formation history, as deduced from the comoving UV intensity as 
a function of redshift.

\begin{figure}[h]
\centerline{\rotatebox{90}{\includegraphics[width=3.5in]{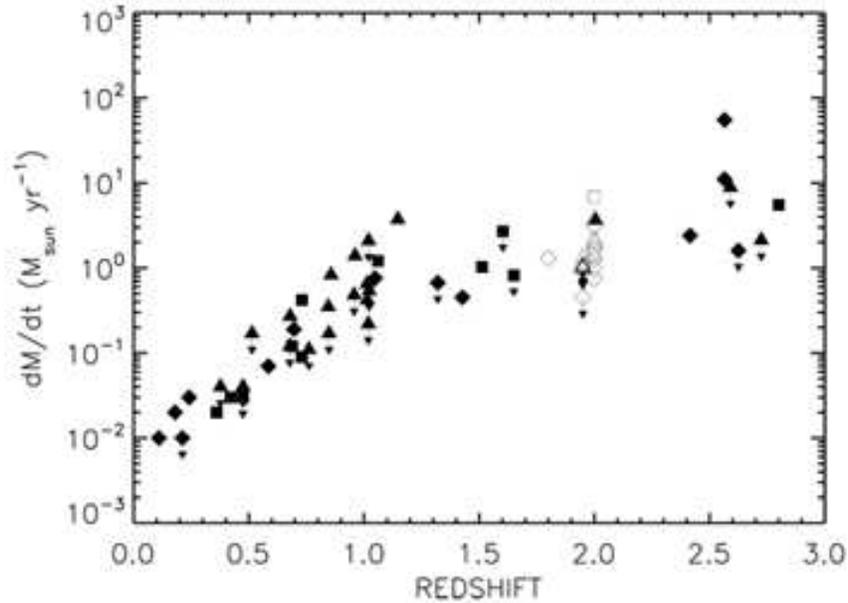}}}
\vskip -0.0in
\caption{The black hole accretion rate, in units of solar masses per year, versus
redshift, assuming a canonical efficiency rate of $10\,\%$ conversion of rest mass
energy into radiation. To generate a $10^9\;M_\odot$ black hole over an accretion
period of order $0.5$ Gyr, a rate of order $2\;M_\odot$ yr$^{-1}$ is required.
(From Barger et al. 2001)}
\vskip 0.0in
\end{figure}

\begin{figure}[h]
\centerline{\rotatebox{90}{\includegraphics[width=3.5in]{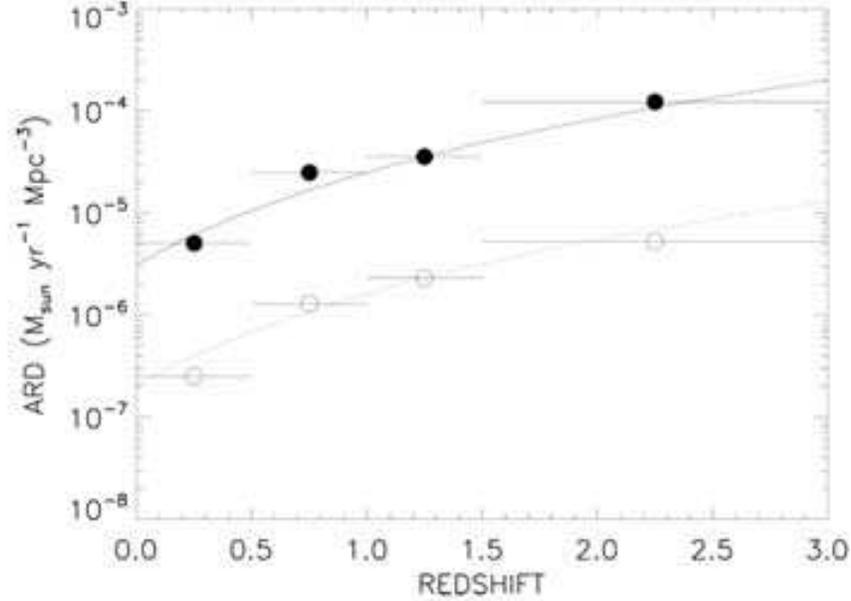}}}
\vskip -0.0in
\caption{The time history of the accretion rate density (effectively, the
rate at which black hole matter is increasing, in units of solar masses per year per megaparsec
cubed), integrated from the data exhibited in figure~12.  The upper (solid) and lower (open)
bounds in this figure correspond to the accretion rate density calculated from the bolometric
and X-ray luminosities, respectively.  The curves illustrate a $(1+z)^3$ dependence, normalized
to the $1<z<1.5$ redshift bin.  (From Barger et al. 2001)}
\vskip 0.0in
\end{figure}

Integrating this curve over time, we infer that about $0.006$ of the all
the mass contained within the bulge of any given galaxy should be in the
form of a central supermassive black hole.  This is close to the value of 
$\sim  0.002$ found in the local neighborhood of our Galaxy.  With these 
facts in hand, and satisfied that most supermassive black holes and their 
host galaxies grew interdependently, one may hypothesize that once created,
a primordial condensation of matter continues to grow with a direct
feedback on its surroundings.  This may happen either because the
quasar heats up its environment and controls the rate at which
additional matter can fall in from its cosmic neighborhood,
or because mergers between galaxies affect the growth of colliding
black holes in the same way that they determine the energy (and
therefore the average speed) of the surrounding stellar distribution.
In support of the idea that massive black holes may have formed
prior to the epoch of galaxy definition, Silk and Rees (1998)
suggest further that protogalactic star formation would have been
influenced significantly by the quasar's extensive energy outflows.
The ensuing feedback on the galaxy's spheroidal component could
be the reason we now see such a tight correlation between the
mass of the central object and the stellar velocities much farther
out.

Another possible reason why black hole masses are correlated with
the velocity dispersion of the host galaxy bulges may be the importance
of galaxy mergers to the hierarchical construction of elaborate elliptical,
or disk-plus-bulge profiles.  Most large galaxies have experienced at 
least one major merger during their lifetime. Larger galaxies have
grown with a succession of collisions and mergers, a process contributing
significantly to the variety of shapes encompassed by the Hubble sequence. 
Meanwhile, the turbulence generated in the core of the collision drives 
most of the gas into the middle, where it forms new stars and feeds
a central black hole, or a pair of black holes. The tight correlation 
between the black-hole mass and the halo velocity dispersion may be 
a direct consequence of this cosmic cascade.

\section{Imaging a Supermassive Black Hole}
Of course, the most exciting development in black hole research would be
the actual imaging of its event horizon (Melia 2003b).  This may be feasible
in the next several years, with the development of a global mm-VLBA. The spectrum 
of Sagittarius A* at the galactic center displays the well-known profile of a 
self-absorbed emitter longward of $3-7$ mm, and a transparent medium at higher 
frequencies (see figure~14). The fact that Sagittarius A* becomes transparent 
across the mm/sub-mm bump in its spectrum means that a depression in intensity
(literally, a `black hole') should arise shortward of $\sim 1$ mm, due to the
effects of strong light bending near the event horizon. The `shadow' cast by the 
black hole in this fashion should have an apparent diameter of about 5 Schwarzschild 
radii (Falcke, Melia, and Agol 2000), nearly independent of the black hole
spin or its orientation. The distance to Sagittarius A* being roughly $8$ kpc, this diameter
corresponds to an angular size of $\sim 30\,\mu$as, which approaches the resolution of
current radio interferometers.

\begin{figure}[h]
\centerline{\includegraphics[width=4in]{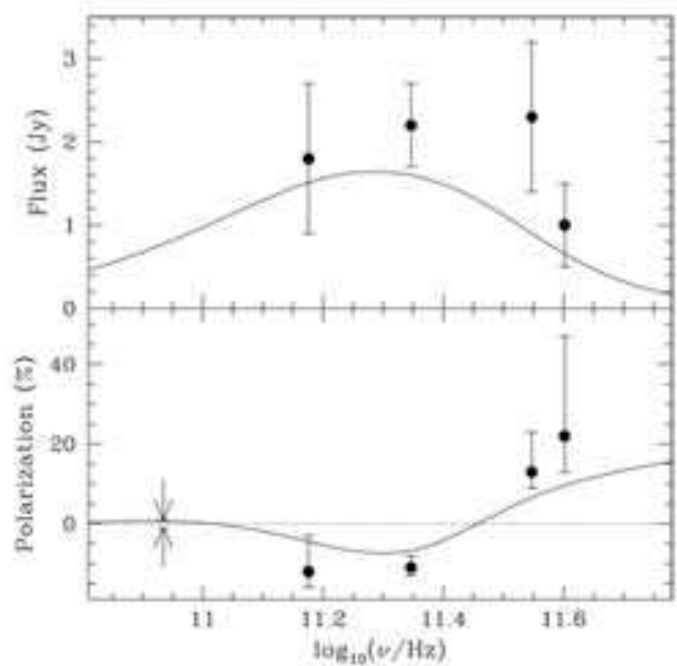}}
\vskip -0.1in
\caption{Spectra showing total flux density and polarization from Sagittarius A*.
The high-frequency data points are from Aitken et al. (2000) and the curves
are from the magnetohydrodynamic model discussed in Bromley, Melia, and Liu  (2001).  The
limit at 84 GHz is from Bower et al. (1999).  At even lower frequencies, the best
fit polarization is consistent with $0$.  In this figure, `negative' polarization
corresponds to the polarization vector being aligned with the spin axis of the black
hole, whereas positive is for a perpendicular configuration.  The polarization
crosses $0$ when this vector flips by $90^\circ$ (Bromley, Melia, and Liu 2001).}
\end{figure}

The fortunate aspect of this transition from optically-thick to optically-thin
emission across the mm/sub-mm bump is that it happens to occur at roughly the 
same frequency where scatter broadening in the interstellar medium decreases 
below the intrinsic source size.  Taking this effect into account, and the finite 
telescope resolution, one infers (Falcke, Melia, and Agol 2000) that the shadow of 
Sagittarius A* should be observable below $\sim 1$ mm. 

Most recently, ray-tracing simulations have taken into account the magnetized 
Keplerian flow and the effects of light-bending and area amplification 
to produce the most detailed theoretical predictions of the images that mm-astronomy 
will produce in the near future.  The polarimetric images shown in figure~15 
demonstrate that future mm/sub-mm interferometry can directly reveal material 
flowing near the horizon in Sagittarius A*.  

\begin{figure}[h]
\centerline{\includegraphics[width=\textwidth]{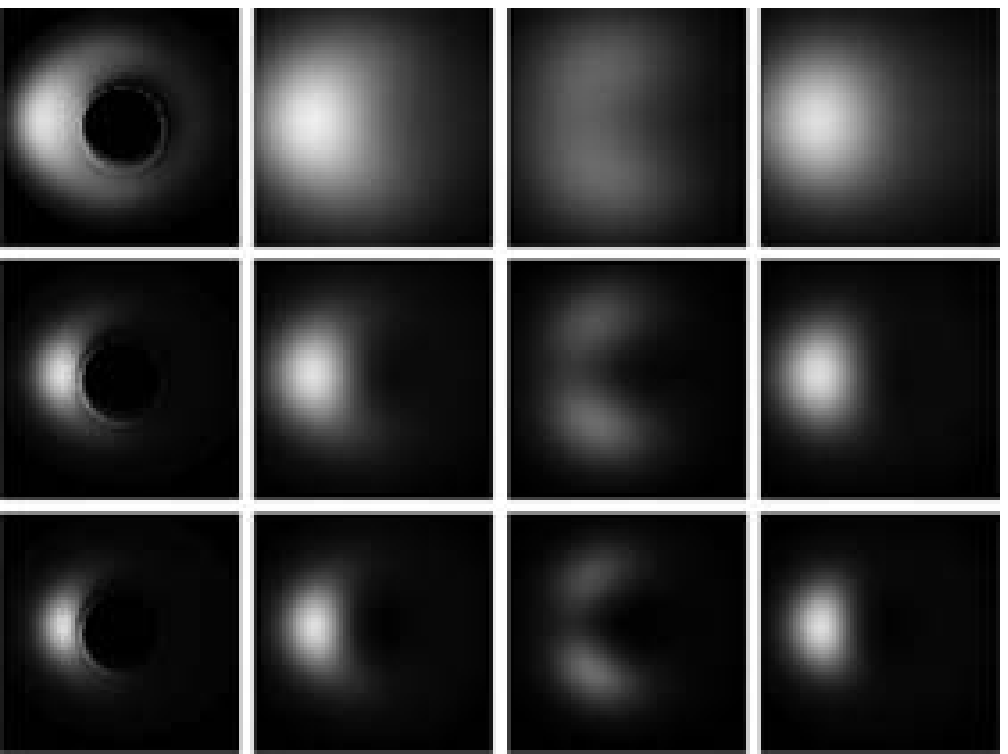}}
\vskip -0.0in
\caption{Polarization maps at three wavelengths near the
peak of the mm to sub-mm emission from Sagittarius A*. The top row shows
emission at 1.5~mm, the middle row is at 1~mm, and the
bottom row corresponds to 0.67~mm. The images in each row
show the raw ray-tracing output (first column on the left), and
an image blurred to account for finite VLBI resolution and
interstellar scattering (second column). The two rightmost
columns give the vertical and horizontal components of the polarized emission.
The pixel brightness in all images scales linearly with flux (Bromley, Melia, and Liu 2001).}
\vskip 0.0in
\end{figure}

The blurring in these images is modeled by convolution with two Gaussians: the first
is an approximate model of the scattering effects with an ellipsoidal filter
whose major- and minor-axis FWHM values are 24.2~$\mu$as~$\times$ $(\lambda/1.3\;\hbox{\rm
mm})^2$ and 12.8~$\mu$as~$\times (\lambda/1.3\;\hbox{\rm mm})^2$, respectively,
for emission at wavelength $\lambda$. In the simulations
shown in Figure~15, the spin axis of the disk was chosen to lie along
the minor axis of the scattering ellipsoid, and a global interferometer
was assumed with a $8,000$ km baseline.  The second is a spherically symmetric 
filter with a FWHM of 33.5~$\mu$as~$\times (\lambda/1.3\;\hbox{\rm mm})$ to 
account for the resolution effects of an ideal interferometer.  These images 
demonstrate clearly the viability of conducting polarimetric imaging of the
black hole at the Galactic center with upcoming VLBI techniques.

\section{Conclusion}
\noindent
We have come a long way since 1963, when Roy Kerr discovered the spacetime
that is now synonymous with black hole research. The past decade, in particular,
has seen a dramatic shift in our view of the role played by supermassive black
holes in the formation of structure in the universe. We now appear to be on the 
verge of finally ``seeing" some of these objects, providing more direct 
evidence for their existence, and a more satisfying confirmation of their 
properties predicted by general relativity.  Within the framework of the Kerr
metric, Sagittarius A* at the Galactic center, and possibly even the central
black hole in the relatively nearby M81 galaxy, are expected to produce distinct
shadows with a precise shape and size of the region where light bending
and capture are important.  This coming decade may finally give us a view into
one of the most important and intriguing predictions of general relativity.

\end{document}